\documentclass[aps,pra,preprint,longbibliography,nofootinbib,10pt]{revtex4-1}

\usepackage{amsmath,amsthm,amsfonts,amssymb,bm}
\usepackage{mdframed,graphicx,array,MnSymbol}
\usepackage[usenames,dvipsnames]{xcolor}
\definecolor{OceanBlue}{rgb}{0,0.35,0.7} 
\usepackage{hyperref}
\hypersetup{colorlinks=true,allcolors=OceanBlue}

\newtheorem{thm}{Theorem}[section] % Theorem
\theoremstyle{definition}
\theoremstyle{remark}
\newmdenv[
  topline=false,
  bottomline=false,
  rightline=false,
  skipabove=\topsep,
  skipbelow=\topsep,
  leftmargin=-5pt,
  rightmargin=-10pt,
  innertopmargin=0pt,
  innerbottommargin=0pt
]{leftrule} 

\newcommand{\pars}[1]{\left(#1\right)} % Parentheses
\newcommand{\bracs}[1]{\left[#1\right]} % Brackets
\newcommand{\mr}{\mathrm} % Text in math mode
\newcommand {\cl}{\mathcal} % Calligraphic math
\DeclareMathOperator{\Tr}{Tr\,}            % Trace
\newcommand*\diff{\mathop{}\!\mathrm{d}}
\newcommand {\ket}[1] {\left|{#1}\right\rangle}
\newcommand {\kets} [1] {\left|#1\right\rangle_{S}}

\newcommand {\keta} [1] {\left|#1\right\rangle_{A}}

\newcommand {\bra}[1] {\langle{#1}|}
\newcommand{\braket}[2]{\langle{#1}|{#2}\rangle}

\newcommand{\abs}[1]{\left | #1 \right |}   % Absolute value
\newcommand{\mean}[1]{\left\langle #1 \right\rangle}  % Angle brackets
\begin{document}
\title{Fundamental quantum limits in optical metrology\\ from rate-distortion theory}
\author{Ranjith Nair}
\affiliation{Department of Electrical \& Computer Engineering,
National University of Singapore,
Singapore 117583}
\date{\today}
\begin{abstract}  We derive fundamental lower bounds on the performance of optical metrology and communication systems  in a Bayesian framework. The derivation uses classical rate-distortion theory in conjunction with bounds on the capacity to transmit classical information of various optical channels specified by the system design. The bounds are expressed in terms of the system parameters, the prior probability distribution of the parameter, and the average energy, i.e., number of photons $E$ in the probe state. For phase estimation, the bounds pertain to a cyclic mean squared error criterion incorporating the cyclic nature of the phase.  In the absence of optical loss, our bounds are applicable to multimode linear phase modulation schemes (including ancilla-assisted ones), and to nonlinear modulations on a single mode. The bounds display inverse-quadratic Heisenberg-limit scaling of the  cyclic mean square error with respect  to $E$. In the presence of any finite amount of loss, a lower bound on ancilla-assisted phase estimation with standard-quantum-limit (SQL)  scaling  is derived, which is shown to be little different from a similar bound for coherent-state probes. For systems involving a single optical mode, we also obtain lower bounds on the mean squared error of estimating any classical parameter, and on the average error probability of any $M$-ary communication system under an average energy constraint. The bounds are valid for arbitrary quantum measurements, for any prior probability distribution, and do not rely on unbiasedness assumptions.
\end{abstract}
\maketitle

\section{Introduction}

Quantum metrology, which may be defined as the science of the fundamental quantum limits on precision measurement using finite resources, has traditionally been studied in the nonrandom (or frequentist) setting in which the classical parameter(s) that are being measured are considered to be unknown but no prior probability distribution is assigned to them. This approach is analogous to that in classical estimation theory, in which the well-known Cram\'er-Rao bound \cite{Cra16,*Rao45,*VanTreesI} lower-bounds the variance of any unbiased estimator of an unknown parameter.  In the quantum domain, the  parameter modulates the quantum state of a physical system and a corresponding quantum Cram\'er-Rao bound \cite{Hel76,*Hol11,*Hay05asymptotic} lower-bounds the variance of any estimate of the parameter obtained from an unbiased quantum measurement.  In this setting, the key quantity is the so-called \emph{quantum Fisher information} (QFI) which is a `local' quantity that measures the degree to which the quantum state varies as the parameter is varied. The  reader is referred to the review articles \cite{GLM11,TA14,D-DJK15,DRC17} on the theory and practice of quantum metrology and sensing in various physical systems using the local QFI-based approach -- the  article \cite{D-DJK15} is devoted specifically to fundamental quantum limits in optical interferometry.

While giving rise to an elegant and rich theory, the QFI-based approach gives valid bounds only for measurements that are unbiased. In principle, the bounds may be achievable in the asymptotic limit \cite{Hay05asymptotic}, although little seems to be known in the way of asymptotic achievability results for energy-constrained estimation schemes. This situation has given rise to many claims on the scaling of the variance  that do not seem to be achieved in reality -- see \cite{HW12b,Hal13} for discussions on this issue. In practice, one is typically faced with a finite number of quantum systems and/or resources with which to estimate the parameter and for which an asymptotic analysis is insufficient. When multiple parameters are to be estimated, the question of achievability of QFI-based bounds becomes even more subtle \cite{Hel76,RJD-D16}.

In contrast to the local approach, the Bayesian approach in statistical estimation assumes a prior probability distribution on the parameter(s) to be estimated and considers the `global' mean squared error (MSE)  of an estimate averaged over this prior distribution. In quantum metrology, this provides an alternative to the QFI-based approach and opens the possibility of obtaining rigorous lower bounds on the MSE of metrology schemes that are  valid in the non-asymptotic regime and without restrictive assumptions on the quantum measurements that can be made. For example, it has long been argued  that the MSE in sensing an optical phase shift can exhibit, at best, an inverse quadratic scaling with the mean number of photons in the quantum state used to sense the phase \cite{YMcCK86,*SM95,*Ou96}. A few years ago, following claims based on the QFI-based approach that this so-called \emph{Heisenberg limit} on phase estimation may be beaten  \cite{ARC+10,RL12,ZJC+13}, several authors revived the subject by providing rigorous proofs of lower bounds -- not limited to optical interferometry -- with Heisenberg-limit scaling \cite{HBZ+12,Tsa12b,GLM12,GM12,HW12a,Nai12ratedist,HW12b,LT16}. These proofs use diverse techniques, e.g., the speed limit on quantum evolutions \cite{GLM12,GM12}, the entropic uncertainty relations \cite{HBZ+12,HW12a},  the quantum Ziv-Zakai bound \cite{Tsa12b,GM12,ZF14},  methods from information theory \cite{Nai12ratedist,HW12b}\footnote{This paper is an updated and extended version of Ref.~\cite{Nai12ratedist}. The results for lossless estimation are closely related to those of \cite{HW12b}.}, and the quantum Weiss-Weinstein bound \cite{LT16}.  

In this paper, we elaborate on a method combining classical rate-distortion theory with quantum information bounds to obtain fundamental limits on parameter estimation and communication in optical systems. The method was pioneered by Yuen \cite{Yue92,Yue04qs} and has been developed in the works \cite{Nai12ratedist,HW12b,Hal18}. While the works \cite{Yue92,Yue04qs,HBZ+12,Tsa12b,GLM12,GM12,HW12a,HW12b,LT16,Hal18} deal exclusively with noiseless estimation scenarios, we show here that the approach can provide strong bounds even in noisy scenarios by applying it to phase estimation in the presence of the ubiquitous optical loss.

Rate-distortion theory is a branch of classical information theory \cite{Sha48a,*Sha48b,CT06} that was introduced in ref.~\cite{Sha48a} and elaborated in ref.~\cite{Sha59,*Sha59bookchapter} by Shannon, and forms the theoretical basis for the lossy compression of data sources. In the simplest scenario involving a continuous data source, the source generates an output modeled as a real-valued random variable $U$, and we wish to map $U$ to another random variable $V$ -- one that perhaps presents lesser storage requirements -- in such a way that a predefined distortion measure  such as the  mean squared error between $U$ and $V$ is kept below a tolerable level $D$. Roughly speaking, rate-distortion theory tells us how much information must remain in $V$ in order to do so, and shows that coded schemes can achieve this compression limit. A fascinating historical introduction into the theory and practice of lossy data compression and other applications of rate-distortion theory may be found in refs.~\cite{BG98,Ber03rd}.

The development of quantum information theory \cite{NC00,Wil17qit} in the past few decades has been much influenced by the ideas of classical information theory, including rate-distortion theory.  One of the first results of quantum information theory, the noiseless coding theorem of Schumacher \cite{Sch95} is a quantum version of Shannon's noiseless source coding theorem, which itself corresponds to the rate-distortion theory with allowed distortion set to zero.  More recently, there have been efforts to formulate a quantum rate-distortion theory applicable to the lossy compression of quantum rather than classical information sources \cite{Bar00,*DB02,*DHW13}.

In this paper, we apply the rate-distortion theory approach to optical quantum metrology under average energy, (i.e., photon number) constraints in the Bayesian setting. In Sec.~\ref{sec:RDtheory}, we review the relevant background from rate-distortion theory and also introduce the cyclic mean squared error measure that is relevant to phase estimation. In Sec.~\ref{sec:ITI}, we review the  theorem from rate-distortion theory  that is basic to our results in this paper -- the so-called Information Transmission Inequality. The application of these tools to quantum metrology begins in Sec.~\ref{sec:1modelimits} where we consider arbitrary estimation and communication schemes involving a single optical mode.  We also argue that for lossless phase estimation, the quantum limit derived in this section applies to multimode ancilla-assisted schemes as well. In Sec.~\ref{sec:mmmp}, we consider multimode multipass protocols for phase estimation and derive a bound on their MSE with Heisenberg-limit scaling. In Sec.~\ref{sec:lossyphaseest}, we obtain a fundamental limit on the MSE of any lossy ancilla-assisted phase estimation scheme exhibiting the inverse-linear \emph{standard quantum limit} (SQL) scaling with the average energy. We conclude with a discussion and outlook in Sec.~\ref{sec:disc}.  
 
\section{Rate-Distortion Theory and the Information Transmission Inequality}

\subsection{Distortion measures, the Rate-Distortion Function, and Shannon's Lower Bound}  \label{sec:RDtheory}

Suppose that we are given a real-valued random variable $X$ with prior probability density $P_X(x)$ \footnote{As is conventional in information theory, we use upper case letters for random variables (though not exclusively for random variables) and the corresponding lower case letters to denote their instances.}. In rate-distortion theory, $X$ is viewed as a data source with differential entropy $h(X)$ given by
\begin{align}\label{DifferentialEntropy}
h(X) = -\int_{-\infty}^{\infty} \diff x\, P_X(x) \ln P_X(x)
\end{align}
and measured in nats/symbol. For another real-valued random variable $\check{X}$ representing a noisy estimate of $X$, we can define the squared error \emph{distortion measure}
\begin{align} \label{sqerror}
d\pars{x,\check{x}} = \pars{\check{x} - x}^2,
\end{align}
and the mean square error (or MSE) as its ensemble average:
\begin{align} \label{MSEdefinition}
{\rm MSE}:= \mathbb{E}d\pars{X,\check{X}} = \iint_{-\infty}^{\infty} \diff x \diff \check{x}\, P_X(x) P_{\check{X}|X}(\check{x}|x)\left (\check{x}- x\right)^2,
\end{align}
where $P_{\check{X}|X}(\check{x}|x)$ is the conditional density of $\check{X}$ given $X$. Here $\mathbb{E}$ denotes statistical expectation of the random function following it.

The  squared error \eqref{sqerror} is but one example of a distortion measure between two real-valued quantities, albeit the most widely used one. Another example is the absolute error $\abs{\check{x} - x}$. For $X$ and $\check{X}$ taking values in the same finite set, a commonly used distortion measure is the so-called Hamming distortion $1 - \delta_{x,\check{x}}$ whose ensemble average is the average error probability $P_e \equiv \mathrm{Pr}\bracs{\check{X} \neq X}$. This distortion measure is considered in Sec.~\ref{sec:1modecomm} in connection with limits on $M$-ary digital communication.  We mention in passing that, in the general theory, the set of values taken by $\check{X}$ need not be the same as the set of values taken by $X$, although we will not consider such cases here.

For the purposes of phase estimation that is the main focus of this paper, the squared error distortion measure must be slightly modified in order to account for the cyclic nature of the phase. Let the random variable $\Phi$ denote the phase to be estimated. We assume it takes values in  $(-\pi, \pi]$ (possibly on only a subset thereof), although any other  interval of length $2\pi$ would do just as well. For  an estimate $\check{\Phi} \in (\pi, \pi]$ of $\Phi$, it is natural to define a \emph{cyclic squared error} distortion measure as
\begin{align} \label{cyclicsqerror}
\overset{\;\circ}{d}({\check{\phi},\phi}) :=
 \left\{
	\begin{array}{ll}
		(\check{\phi} - \phi)^2  & \mbox{if } \abs{\check{\phi} -\phi} \leq \pi,  \\
		\bracs{2\pi -  \abs{\check{\phi} -\phi}}^2  & \mbox{otherwise.}
	\end{array}
\right.
\end{align}
In other words, $\overset{\;\circ}{d}({\check{\phi},\phi}) $ is the squared length of the \emph{shorter} of the two arcs between $\check{\phi}$ and $\phi$ on the unit-radius phase circle. The resulting \emph{cyclic mean squared error} is then
\begin{align} \label{CMSEdef}
\mathrm{CMSE} := \mathbb{E}\overset{\;\circ}{d}\pars{{\check{\Phi},\Phi}}.
\end{align}
If one maps the phase circle onto any interval of length $2\pi$ on the real line, we clearly have $ \overset{\;\circ}{d}({\check{\phi},\phi}) \leq d({\check{\phi},\phi})$, where the latter quantity is the squared error distortion \eqref{sqerror} between $\phi$ and $\check{\phi}$ considered simply as real numbers in the chosen $2\pi$-interval ignoring their cyclic nature. 
 
For an arbitrary source $X$ and distortion measure $d(\check{x},x)$, the  \emph{rate-distortion function} $R(D)$ is defined as \cite{Sha59,CT06}
\begin{align} \label{R(D)}
R(D) = \inf_{P_{\check{X}|X}(\check{x}|x): \mathbb{E}d(X,\check{X}) \leq D} I(X; \check{X}),
\end{align}
where the quantity being minimized is the mutual information $I(X;\check{X})$ between the source and estimate and the infimum is over all conditional distributions $P_{\check{X}|X}(\check{x}|x)$ that yield average distortion less than or equal to $D$. Note that $R(D)$ depends on the prior distribution $P_X(x)$ as well as the given distortion measure. 

The rate-distortion function $R(D)$ may be thought of informally as the amount of non-redundant information per symbol emitted by the source when allowing for an average distortion of up to $D$ in a reconstructed version of the source. Examples of the computation of $R(D)$ for some standard sources and distortion measures may be found in \cite{Sha59,CT06}, although numerical evaluation is usually required for an arbitrary source. In general, the function $R(D)$ and its inverse $D(R)$ are decreasing and convex in their respective arguments. For a real-valued random variable $X$ and the squared error distortion measure \eqref{sqerror}, the following lower bound on the rate-distortion function (called the \emph{Shannon lower bound}) has been derived \cite{Sha59,BG98,CT06}:-
\begin{align} \label{R(D)LB}
R(D) \geq \frac{1}{2} \ln\left(\frac {Q_X}{D}\right),
\end{align}
where 
\begin{align}
Q_X = \frac{1}{2 \pi e}\, e^{2h(X)},
\end{align} 
is the \emph{entropy power} of $X$ \cite{CT06}. Note that the bound is useful for $D \in (0,Q_X]$ (outside which it can be replaced by zero) and is convex and decreasing on this interval. The lower bound 
\begin{align} \label{D(R)LB}
D(R) \geq Q_X \,e^{-2R}
\end{align}
on $D(R)$ follows from eq.~\eqref{R(D)LB}.

For a phase random variable $\Phi$, an estimate $\check{\Phi}$ thereof, and the cyclic squared error measure, we have the rate-distortion function
\begin{align} \label{R(D)cysqerror}
R(D) := \inf_{P_{\check{\Phi}|\Phi}(\check{\phi}|\phi): \mathbb{E}\overset{\;\circ}{d}({\check{\Phi},\Phi}) \leq D} I\pars{\Phi;\check{\Phi}}.
\end{align}
We now derive a Shannon lower bound on \eqref{R(D)cysqerror} that will be needed for our later results. First note that we may write the quantity to be minimized in Eq.~\eqref{R(D)cysqerror} as
\begin{align}
I\pars{\Phi; \check{\Phi}} &= h(\Phi) - h\pars{\Phi | \check{\Phi}}  \label{eq1}\\
&=  h(\Phi) - \int_{-\pi}^{\pi} \diff \check{\phi}\, P_{\check{\Phi}}\pars{\check{\phi}} h\pars{\Phi | \check{\Phi} = \check{\phi}},
\end{align}
where $P_{\check{\Phi}}\pars{\check{\phi}}$ is the marginal distribution of $\check{\Phi}$ and
\begin{align}
h\pars{\Phi | \check{\Phi} = \check{\phi}} = - \int_{-\pi}^{\pi} \diff \phi \, P_{\Phi|\check{\Phi}} \pars{\phi | \check{\phi}}\, \ln\, P_{\Phi|\check{\Phi}}\pars{\phi | \check{\phi}},
\end{align}
i.e., the differential entropy of the conditional distribution of $\Phi$ on the phase circle for a given value of $\check{\phi}$. We now upper bound this quantity as follows. Suppose the phase circle is cut and unrolled into an interval of the real line centered at $\check{\phi}$ and that $\Phi$ is now considered as a real-valued random variable. Such an operation leaves the differential entropy unchanged. Further, for any value $\phi$ of $\Phi$ the cyclic squared error $\overset{\,\circ}{d}(\check{\phi},\phi)$ agrees with the squared error $\pars{\phi - \check{\phi}}^2$. Now the expectation (over the conditional distribution of $\Phi$ alone) $\mathbb{E}_{\Phi} \pars{\Phi - \check{\phi}}^2 \geq \mbox{Var}[\Phi|\check{\phi}]$, the conditional variance of $\Phi$ given $\check{\phi}$. Given the fact that of all probability distributions on the real line with variance $V$, the Gaussian distribution has the maximum differential entropy $\frac{1}{2} \ln (2\pi e V)$ \cite{CT06}, we have
\begin{align}
h\pars{\Phi | \check{\Phi} = \check{\phi}}
&= \int_{-\pi}^{\pi} \diff \check{\phi}\, P_{\check{\Phi}}\pars{\check{\phi}} h\pars{\Phi | \check{\Phi} = \check{\phi}} \\
& \leq \frac{1}{2} \int_{-\pi}^{\pi} \diff \check{\phi}\, P_{\check{\Phi}}\pars{\check{\phi}} \ln \pars{2\pi e\, \mathbb{E}_{\Phi} \pars{\Phi - \check{\phi}}^2} \\
&= \frac{1}{2} \int_{-\pi}^{\pi} \diff \check{\phi}\, P_{\check{\Phi}}\pars{\check{\phi}} \ln \pars{2\pi e\, \mathbb{E}_{\Phi}\overset{\,\circ}{d}\pars{\Phi, \check{\phi}}} \\
& \leq  \frac{1}{2}  \ln \pars{2\pi e\,  \int_{-\pi}^{\pi} \diff \check{\phi}\, P_{\check{\Phi}}\pars{\check{\phi}} \mathbb{E}_{\Phi}\overset{\,\circ}{d}\pars{\Phi, \check{\phi}}} \\
&=  \frac{1}{2}  \ln \pars{2\pi e\,  \mathbb{E}\overset{\,\circ}{d}\pars{\Phi, \check{\Phi}}}\\
&\leq  \frac{1}{2}  \ln \pars{2\pi e D}.
\end{align}
Here, we have used the concavity of the logarithm and the fact that the maximum allowed CMSE in the minimization of Eq.~\eqref{R(D)cysqerror} is $D$. Together with \eqref{eq1}, we thus have the lower bounds
\begin{align} \label{RDSLB}
R(D) \geq \frac{1}{2} \ln \pars{\frac{Q_\Phi}{D}} \equiv \underline{R}(D),
\end{align}
and
\begin{align} \label{DRSLB}
D(R) \geq Q_\Phi \,e^{-2R} \equiv \underline{D}(R),
\end{align}
where, $Q_\Phi = e^{2h(\Phi)}/(2\pi e)$ is the entropy power of $\Phi$. While these bounds are identical in form to the lower bounds \eqref{R(D)LB} and \eqref{D(R)LB} for the squared error measure, it is important to note that the average distortion appearing in the bounds \eqref{RDSLB} and \eqref{DRSLB} is the average of the \emph{cyclic} squared error distortion and the latter bounds do not follow as special cases of the former.

\subsection{The Information Transmission Inequality} \label{sec:ITI}

\begin{figure}   \begin{center}
   \begin{tabular}{c}
   \includegraphics[trim=40mm 40mm 44mm 30mm, clip=true,width=0.9\columnwidth]{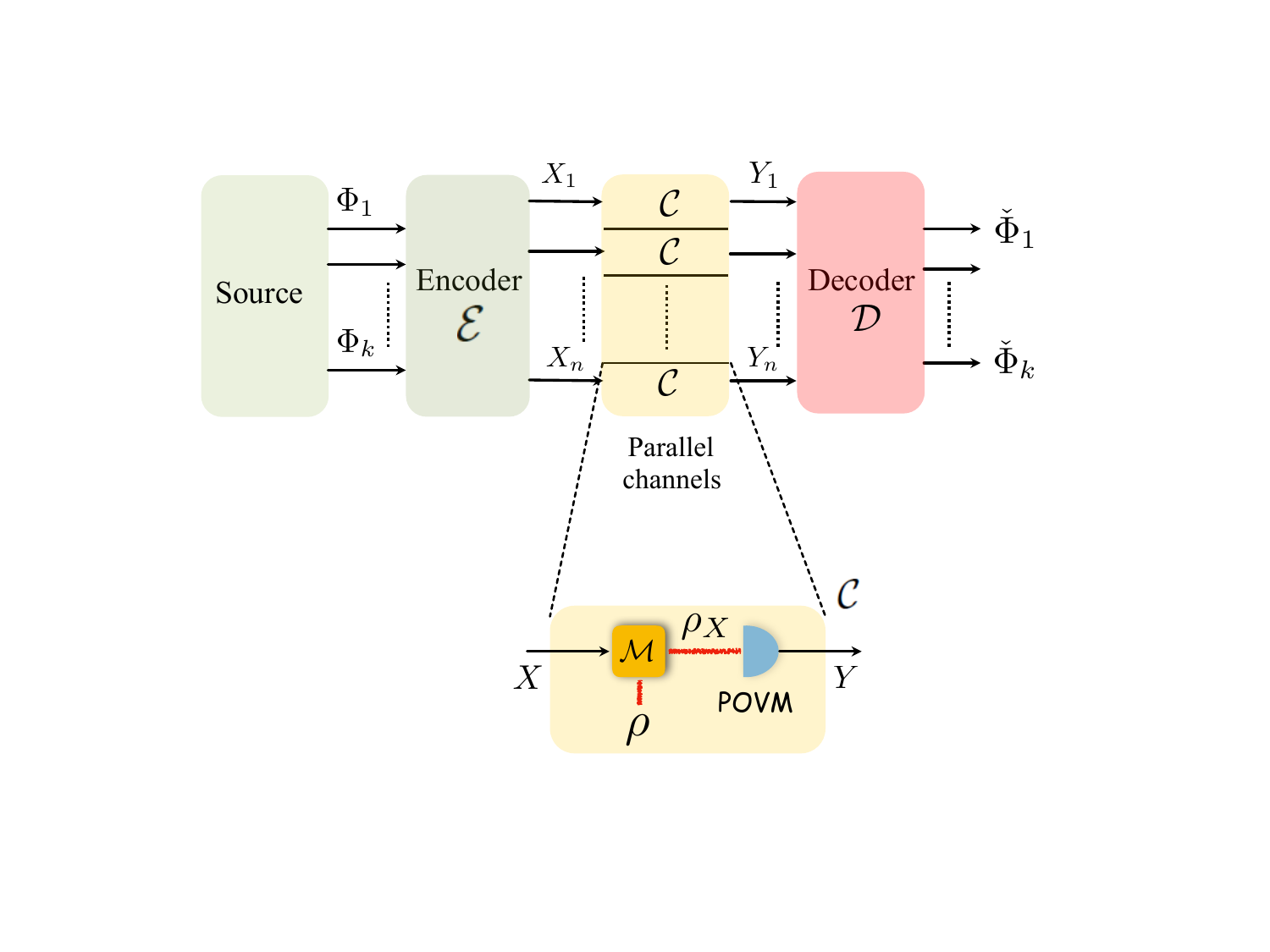}
   \end{tabular}
   \end{center}
   \caption{Block diagram illustrating the parallel encoding, channel transmission, and decoding operations on a sequence of source ouputs to which the Information Transmission Inequality applies. In its application to quantum metrology, each of the parallel channels $\mathcal{C}$ is realized by a modulation of $X$ on to density operators of a quantum system followed by a quantum measurement (POVM) on the system (see the blow-up). Thick red lines indicate transmission of quantum systems, which are taken to be optical beams of one or more modes in this paper.} \label{fig:sensingscheme}

\end{figure}

The operational significance of the rate-distortion function is elucidated by the positive and converse parts of the noisy source coding theorems of Shannon \cite{Sha59}. For our purpose of obtaining lower bounds on the achievable average distortion,  the converse part is of primary relevance. The fundamental result in this connection, called the \emph{Information Transmission Inequality} (ITI in the sequel) \cite{BG98}, is stated below (Refer Fig.~\ref{fig:sensingscheme}).

\begin{thm} \label{thm:iti}[\textrm{Information Transmission Inequality -- Theorem 1 of ref.~\cite{Sha59}}] Given $k$ independent and identically distributed (i.i.d.) source outputs $\mathbf{\Phi}=\Phi_1,\ldots,\Phi_k$, each with prior distribution $P_\Phi(\phi)$. For any given distortion measure $d(\Phi,\check{\Phi})$, let the rate-distortion function of the source be $R(D)$ nats/symbol. We are given an encoder $\mathcal{E}$ that maps $\mathbf{\Phi}=\Phi_1,\ldots,\Phi_k$ to an $n$-symbol-long codeword $\mathbf{X}=X_1, \ldots, X_n$ that is transmitted over a channel  $\mathcal{C}$ with capacity $C$ nats/use. Let the channel output codeword be $\mathbf{Y}=Y_1,\ldots, Y_n$ which is mapped by a decoder $\mathcal{D}$ to an estimate $\mathbf{\check{\Phi}}= \check{\Phi}_1,\ldots, \check{\Phi}_k$ of $\mathbf{\Phi}$. Defining a per-symbol average distortion measure $\mathbb{E}{d}(\mathbf{\Phi},\mathbf{\check{\Phi}}) = \sum_{i=1}^k \mathbb{E}d(\Phi_i,\check{\Phi}_i)/k$, we have
\begin{align}\label{ITI}
\mathbb{E}{d}(\mathbf{\Phi},\mathbf{\check{\Phi}}) \geq  D\left(\frac{n}{k}\, C\right),
\end{align}
where $D(\cdot)$ is the function inverse to $R(D)$.
\end{thm}

Note that, unlike the positive part of the noisy source coding theorem which applies in the asymptotic limit of long codes with $n \rightarrow \infty$, Theorem~\ref{thm:iti} applies to any given  system of the form of Fig.~\ref{fig:sensingscheme}.  It has been used to obtain performance lower bounds in classical estimation and communication systems \cite{VanTreesII}. In its application to quantum metrology \cite{Yue92,Yue04qs}, the key point is to implement the classical channel $\mathcal{C}$ appearing in Fig.~\ref{fig:sensingscheme} using a quantum system in the following way.  Given a codeword symbol $X$, we implement a \emph{modulation} map $\mathcal{M}$ that takes the symbol $X$ and a given \emph{probe state} $\rho$ in the Hilbert space $\cl{H}$ of the quantum system of interest into another density operator $\rho_X$ on  $\cl{H}$. We then make a \emph{quantum measurement} on $\rho_X$ described by a Positive-Operator-Valued Measure (POVM) $\{\Pi_Y\}$ \cite{Hol11,NC00}, whose outcome $Y$ is the output codeword symbol (see blow-up of $\mathcal{C}$ in Fig.~\ref{fig:sensingscheme}). Any such choice of probe state, modulation map, and POVM  induces a  probability transition matrix $P_{Y|X}(y|x)$, i.e., a classical channel, for which a channel capacity $C$ may be defined. This $C$ may then be used in eq.~\eqref{ITI} to yield a lower bound on the distortion. The calculation of $C$ can be made to incorporate any constraints relevant to the sensing problem, e.g., an energy constraint on the probe state, a constraint on the kind of modulation allowed, or a constraint on the measurement POVM.

In the context of quantum optics, we can consider a fixed class of probe states, e.g., coherent or quadrature squeezed states, certain kinds of modulation such as phase modulation or displacement in phase space, or restrict ourselves to standard measurements such as photon counting,  homodyne or heterodyne detection. The channel capacities under a mean energy constraint under these probe, modulation, and measurement choices are known in many cases \cite{Yue04qs,Sha09}. Using this approach, performance bounds for the communication or sensing of a Gaussian source were obtained in ref.~\cite{Yue04qs}. In addition, for lossless estimation of a uniform phase parameter, a lower bound exhibiting SQL scaling  was obtained for coherent-state probes, and a lower bound exhibiting Heisenberg-limit scaling was obtained for a quadrature-squeezed-state (or two-photon coherent state (TCS)) probe.
 
In this paper, we are mainly concerned with lower bounds on the CMSE for lossless and lossy phase estimation under a mean energy constraint $E$ on the probe state $\rho$ used to sense the phase.  We will not consider coding over multiple instances of the phase, i.e., we set $k=n=1$ in Fig.~1 so that  $\mathbf{X}=\mathbf{\Phi}\equiv \Phi$ and  $\mathbf{Y}=\mathbf{\check{\Phi}}\equiv \check{\Phi}$. This assumption of no coding is realistic in the single-parameter estimation problem considered here, though it may be relaxed in more general situations.  In line with the  remarks above, a large part of our work consists in estimating the classical capacity of the particular channel resulting from phase modulation of the probe state, while allowing arbitrary POVM measurements on the modulated states. 

\section{Quantum Limits on single-mode phase estimation and communication} \label{sec:1modelimits}

\begin{figure}[tbp]
\begin{minipage}[t]{0.45\linewidth}
    \includegraphics[trim=55mm 80mm 80mm 95mm, clip=true,width=0.8\linewidth]{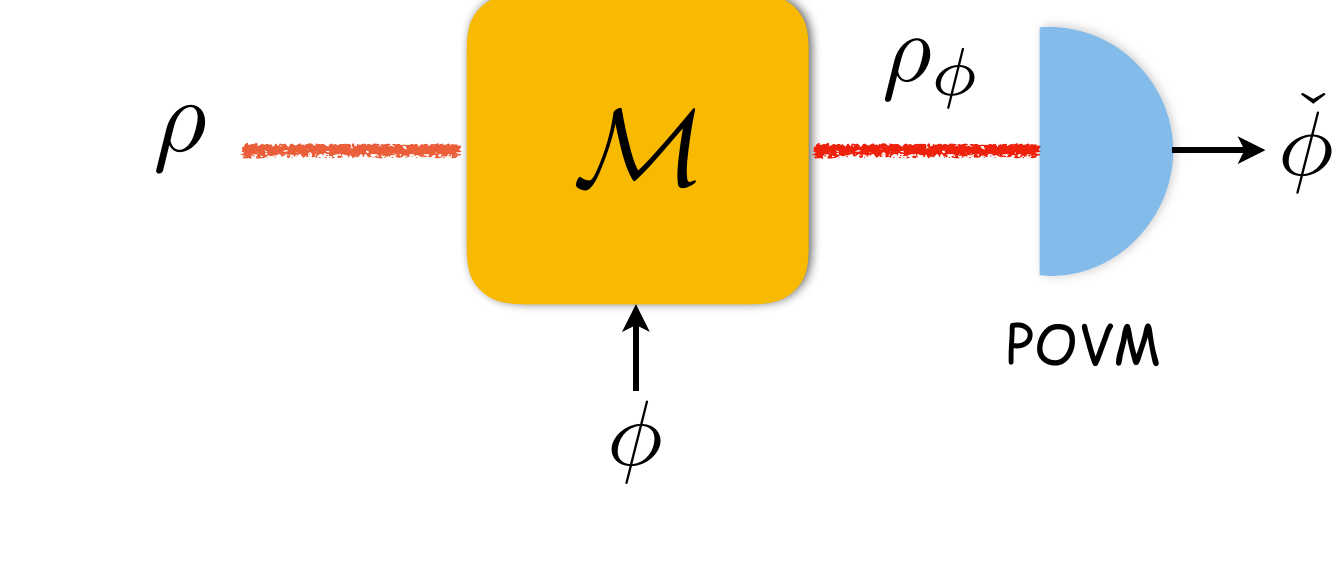}
    \caption{Schematic of a single-mode phase estimation scheme: A phase $\phi$ is modulated on to a probe state $\rho$ satisfying the average energy constraint of Eq.~\eqref{ec}, giving rise to an ensemble of states $\{\rho_\phi\}$ with probability distribution $P_\Phi(\phi)$. An arbitrary POVM is measured to yield an estimate $\check{\phi}$.}     \label{fig:1modephaseest}
\end{minipage}%
    \hfill%
\begin{minipage}[t]{0.45\linewidth}
    \includegraphics[trim=83mm 85mm 75mm 100mm, clip=true,width=0.8\linewidth]{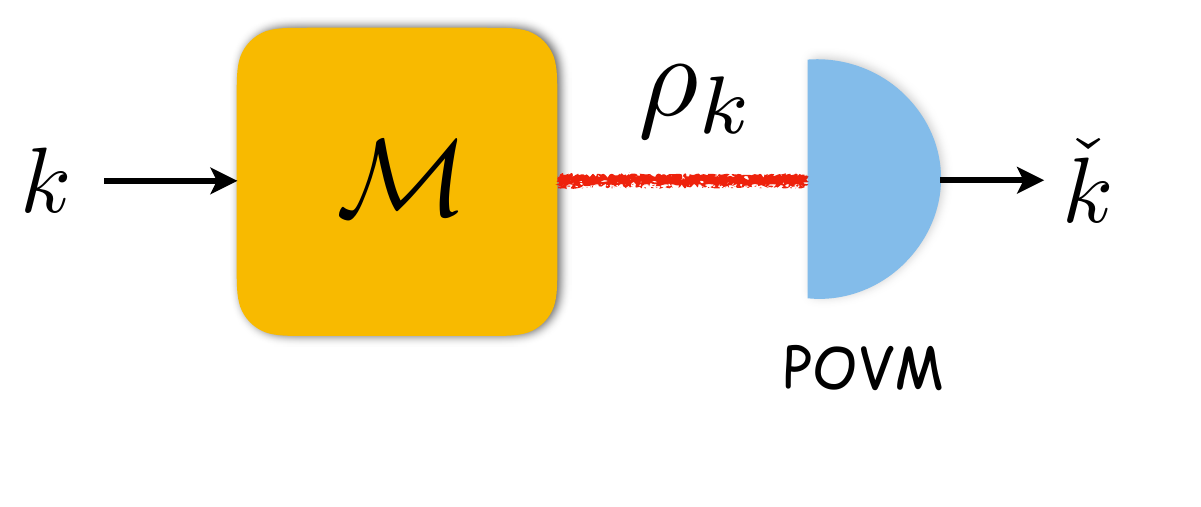}
    \caption{A general single-mode $M$-ary communication scheme: Each of $M$ messages $k \in \{1,\ldots, M\}$ drawn from a probability distribution $\{p_k\}_{k=1}^M$ is mapped to a state $\rho_k$ of a single optical mode such that the output ensemble satisfies the average energy constraint \eqref{commenergyconstraint}. Measuring the state using an arbitrary POVM generates an estimate $\check{k} \in \{1,\ldots, M\}$ of the message.}       \label{fig:1modecomm}
\end{minipage} 
\end{figure}

In this Section, we begin our study of quantum limits by considering metrology and communication scenarios  involving a single optical mode with Hilbert space $\cl{H}$. Fig.~\ref{fig:1modephaseest} depicts an arbitrary phase modulation scheme wherein a phase parameter $\phi$ is modulated onto a probe state $\rho$ of $\cl{H}$ via a Hamiltonian that is an arbitrary function $\hat{G} = f(\hat{N})$ of the number operator $\hat{N}$ of the mode. While $\hat{G} = \hat{N}$ corresponds to the usual linear modulation, many authors have suggested that nonlinear Hamiltonians improve the sensitivity of phase estimation beyond the Heisenberg limit in the local approach based on the QFI \cite{Lui04,BL05,Ou12}. For any such $\hat{G}$, the resulting output ensemble $\cl{E} = \{P_\Phi(\phi), \rho_\phi\}$, where $\rho_\phi = \exp(-i\phi\hat{G}) \rho \exp(i \phi \hat{G})$. The average energy of the probe (and thus each state in the output ensemble) is assumed to be bounded as:
\begin{align} \label{ec}
 \Tr\rho\, \hat{N} \leq E.
\end{align}
For an arbitrary estimator $\check{\Phi}$ of $\Phi$ that results from implementing a  POVM on the ensemble, we are interested in lower bounds on the CMSE of Eq.~\eqref{CMSEdef}. 

Figure~\ref{fig:1modecomm} depicts a single-mode $M$-ary digital communication scheme. Here, an ensemble $\cl{E} = \{p_k, \rho_k\}_{k=1}^M$ consisting of $M$ states $\{ \rho_k \}_{k=1}^M$ of $\cl{H}$ drawn with probabilities $\{p_k\}_{k=1}^M$ is given under the average energy constraint $\sum_k p_k \Tr \rho_k \hat{N} \leq E$.
The sender draws one of the states of the ensemble while the receiver implements a POVM $\{ \hat{\Pi}_{\check{k}} \}_{\check{k}=1}^M$ of his choice that yields an estimate $\check{k}$ of the message chosen by the sender with a view to minimizing the average error probability
\begin{align} \label{PE}
P_e = 1 - \sum_{k=1}^M p_k \Tr \rho_k \hat{\Pi}_k.
\end{align}
In Sec.~\ref{sec:1modecomm}, we will derive a lower bound on $P_e$ in terms of $E$ and $M$.

\subsection{Heisenberg Limits for Noiseless Phase Estimation} \label{sec:1modeestimation}
Consider the single-mode phase estimation scheme of Fig.~\ref{fig:1modephaseest}. As mentioned above the state of the output ensemble corresponding to the value $\phi$ is $\rho_\phi = \exp(-i\phi\hat{G}) \rho \exp(i \phi \hat{G})$ with $\hat{G} = f (\hat{N})$. If the average energy of the probe $\rho$ is less than or equal to $E$, the average energy of the output ensemble is also bounded by $E$. The unrestricted classical capacity $C(E)$ (in nats/use) of a single-mode noiseless channel under a mean energy constraint on the output ensemble is well known \cite{YO93} and is given by
\begin{align} \label{capacity}
g(E)  \equiv (E+1) \ln \left(E +1\right) -E \ln E,
\end{align}
which is an increasing function of $E$.  By `unrestricted', we mean that there is no constraint on the output ensemble of the modulation map other than that the  ensemble has mean energy $E$, and that the POVM used on the output ensemble is arbitrary. In particular, the energy-constrained capacity $C_{\mr{ph}}(E)$ of the phase modulation scheme of Fig.~\ref{fig:1modephaseest} satisfies $C_{\mr{ph}}(E) \leq g(E)$. We can now apply the ITI \eqref{ITI} by taking the channel $\mathcal{C}$ of Fig.~\ref{fig:sensingscheme} to be realized by phase modulation as in Fig.~\ref{fig:1modephaseest} and taking $k=n=1$, $X =\Phi$, and $\check{\Phi}=Y$. Adopting the cyclic square error distortion measure \eqref{cyclicsqerror}, we apply the Shannon lower bound \eqref{DRSLB} to get  (recall that the inverse of the rate-distortion function $D(\cdot)$  is a decreasing function of its argument):
\begin{align}
{\rm CMSE} &\geq {D}(C_{\mathrm{ph}}(E)) \geq D\pars{g(E)} \nonumber\\ 
 &\geq \underline{D}\pars{g(E)}  = Q_\Phi \left(1 + \frac{1}{E}\right)^{-2E}\frac {1}{(E+1)^2} \nonumber\\
& \geq \frac{Q_\Phi}{e^2} \frac {1}{(E+1)^2}. \label{Hlimit}
\end{align}
This lower bound on the CMSE is valid for arbitrary generators $\hat{G}=f\pars{\hat{N}}$ and has the form of a Heisenberg limit by virtue of its inverse quadratic dependence on $E$. Further, it depends on the prior distribution of the phase through its entropy power $Q_\Phi$. For linear modulation $\hat{G} =\hat{N}$, the above argument for Heisenberg-limit scaling was, in essence, given by Yuen in ref.~\cite{Yue92}, though specialized to a uniform prior and for the squared error distortion measure. It is also interesting to note that the above bound agrees with a different Bayesian bound of Hall and Wiseman \cite{HW12a} (see Eq.~(17) and the surrounding discussion therein) for a phase distribution uniformly distributed on an interval of length $L$ in $(-\pi, \pi]$, in which case $Q_\Phi =  L^2/2\pi e$. It also agrees with another bound valid for a uniform prior distribution on $(-\pi, \pi]$ and derived using entropic uncertainty relations \cite{HBZ+12}.

For the important special case of linear phase estimation for which $f(\hat{N}) = \hat{N}$, consider the potentially more powerful   \emph{ancilla-assisted parallel strategy} in which a joint probe state $\rho$ of  $M$ `signal' modes ($S$) and an arbitrary `ancilla' system $A$ with a total average energy $E$ in the signal modes is prepared.  The signal modes each undergo a phase shift $\phi$ while the ancilla modes remain unaffected so that we have the output state
\begin{align} \label{phasemodulation}
\rho_\phi = \hat{U}_\phi\, \rho\, \hat{U}_\phi^\dag 
\end{align}
for 
\begin{align}
\hat{U}_\phi = \left(\bigotimes_{m=1}^M e^{-i \phi \hat{N}_m}\right) \bigotimes \hat{I}_A,
\end{align}
where $\{\hat{N}_m\}_{m=1}^M$ are the photon number operators of the $M$ signal modes and $\hat{I}_A$ is the identity operator on the ancilla system. Finally, a POVM is  implemented on the joint signal-ancilla system to yield an estimate $\check{\Phi}$ of $\Phi$. However, it can be shown \cite{NY11} (see also \cite{NYG+12})\footnote{A sketch of the argument: Without loss of generality, we may consider a pure-state probe of the signal modes and ancilla. We then show that the inner product between the modulated states for any pair of phases $(\phi,\phi')$ is the same as that of a suitably chosen single-signal-mode probe of the same energy. This probe and the original probe must therefore yield the same optimal performance.} that any such strategy has the same performance as a suitably chosen single-mode probe of average energy $E$ or less  so that the same Heisenberg limit \eqref{Hlimit} applies to such strategies. Similarly, it is easy to verify that for any nonlinear generator $f\pars{\hat{N}}$ acting on a single signal mode, any ancilla-assisted scheme is equivalent to a corresponding scheme involving states of the signal mode alone. Such schemes are accordingly subject to the Heisenberg limit \eqref{Hlimit} as well.
%For nonlinear generators, it has same result has been shown using a different approach in Ref.~\cite{HW12b}.

In another direction, we may wonder if the bound \eqref{Hlimit} may be strengthened by evaluating and using in the derivation the capacity $C_{\rm ph}(E)$ of a single-mode channel that is restricted to employ phase modulation on a probe rather than the unrestricted capacity $g(E)$. Indeed, the ensembles  well-known to achieve $g(E)$ are the number states with a thermal distribution \cite{YO93} and coherent states with a circularly-symmetric Gaussian distribution on the phase space \cite{GGL+04}. However, using uniform phase modulation on the  probe state
\begin{align} \label{coherentthermalstate}
\ket{\psi} = \frac{1}{\sqrt{E+1}}\sum_{n=0}^{\infty} \pars{\frac{E}{E+1}}^{n/2}\ket{n},
\end{align}
achieves $C_{\rm ph}(E) = g(E)$  \cite{GDN+11}, so that \eqref{Hlimit} cannot be improved in this way. However, if the probe state is restricted to be a single-mode coherent state $\ket{\sqrt{E}}$, the capacity achievable using any linear or nonlinear Hamiltonian for phase modulation is bounded by the Holevo bound \cite{Hol73a,YO93,NC00} as:
\begin{align}
C_{\rm ph}^{\rm CS}(E) & \leq S \pars{ \int \diff \phi\, P_{\Phi}(\phi) \rho_{\phi}}  \equiv S(\overline{\rho})\\
& \leq S \pars{\cl{N}\pars{\overline{\rho}}},
\end{align}
where $S(\cdot)$ is the von Neumann entropy, $\mathcal{N}$ is the quantum channel given by
\begin{align} \label{Ndef}
\cl{N}(\sigma) := \sum_{n=0}^{\infty} \bra{n}\sigma\ket{n} \ket{n}\bra{n}
\end{align}
and representing a measurement in the photon number basis $\{\ket{n}\}$, and the last inequality follows from the fact that this channel is unital and hence entropy non-decreasing \cite{NC00}. Thus,  $C_{\rm ph}^{\rm CS}(E)$ is upper-bounded by the (Shannon) entropy of a Poisson distribution of mean $E$. The Shannon entropy of an integer-valued random variable $X$ can be bounded in terms of its variance as \cite{Mas88}:
\begin{align} \label{diffentbound}
H(X) \leq \frac{1}{2} \ln\bracs{ 2 \pi e \pars{{\rm Var}\,X + \frac{1}{12}} },
\end{align}
resulting in 
\begin{align}
C_{\rm ph}^{\rm CS}(E) & \leq \frac{1}{2} \ln\bracs{ 2 \pi e \pars{E + \frac{1}{12}}}.
\end{align}
Inserting this bound into the ITI as above results in the lower bound
\begin{align} \label{CSbound}
{\rm{CMSE}} \geq \frac{Q_{\Phi}}{2\pi e \pars{E + \frac{1}{12}}}
\end{align}
which shows an SQL scaling of the CMSE with $E$ (see also the discussion in ref.~\cite{HW12b}).

We note that several authors have derived Bayesian bounds with Heisenberg-limit scaling in terms of the mean value of the \emph{generator} $\hat{G} = f(\hat{N})$ of the modulation \cite{GM12,HW12a,Tsa12b,LT16}. The bounds derived here are rather in terms of the physically relevant probe energy and are stronger than the former for superlinear functions $f$.

Finally, we note that our use of the unrestricted capacity $g(E)$ in the derivation of the Heisenberg limit \eqref{Hlimit} implies that it also applies to \emph{any} modulation scheme of a real-valued parameter $x \in \mathcal{X} \subset \mathbb{R}$ on to states $\rho_x$ of $\cl{H}$ (with the CMSE replaced by the usual MSE \eqref{MSEdefinition}), which may  be neither unitary nor energy-conserving as long as the average energy of the output ensemble is less than or equal to $E$ . This generality is one of the useful features of the rate-distortion approach to Bayesian metrology bounds.

\subsection{Limit on  optical $M$-ary communication} \label{sec:1modecomm}

Consider now the general  $M$-ary digital communication scheme shown in Figure~\ref{fig:1modecomm}. A sender wishes to communicate one of $M$ messages  $\{1,\ldots, M\}$ to a receiver over a single-mode optical channel. The message, represented by the random variable $K$, has the prior probability distribution $\{p_k\}_{k=1}^M$. The modulation map $\cl{M}$ takes message $k$ into a state $\rho_k$ of $\cl{H}$ in any manner that  satisfies the average energy constraint 
\begin{align} \label{commenergyconstraint}
\sum_k p_k \Tr \rho_k \hat{N} \leq E.
\end{align}
The receiver implements a POVM $\{ \hat{\Pi}_{\check{k}} \}_{\check{k}=1}^M$ that yields an estimate $\check{K}$ of the message chosen by the sender with a view to minimizing the average error probability
\begin{align} \label{PE}
P_e &= 1 - \sum_{k=1}^M p_k \Tr \rho_k \hat{\Pi}_k\\
	&= \mathbb{E} d(K,\check{K})
\end{align}
in terms of the Hamming distortion function $d(k, \check{k}) = 1 - \delta_{k,\check{k}}$. Similar to our approach to obtain MSE (or CMSE) bounds for estimation problems, we may consider the rate-distortion function for $K$ with the Hamming distortion measure \cite{CT06,Yeu12first}, derive a Shannon lower bound on it, and apply the ITI along with the unrestricted capacity $g(E)$ of a single-mode channel to get a lower bound on $P_e$. It turns out that the final result is equivalent to applying Fano's inequality \cite{Fan61,CT06} which states in the present context that
\begin{align} \label{Fano}
H_2(K | \hat{K}) \leq h_2(P_e) + P_e\,\log_2(M-1).
\end{align}
Here $h_2(x) = -x\log_2(x) - (1-x)\log_2(1-x)$ is the binary entropy function and the subscripts on the entropies indicate that logarithms to base 2 are being taken. Applying the inequalities  $P_e \leq \sqrt{P_e}$ and $h_2(x) \leq 2\sqrt{x(1-x)} \leq 2\sqrt{x}$ for $0 \leq x \leq 1$ \cite{FvdG99}, and using the Holevo bound on $I(K;\check{K})$, we have $H(K|\check{K}) = H(K) - I(K;\check{K}) \geq H(K) - g(E) \geq H(K) - \ln(1 +E) - 1$. Rearranging, we get
\begin{align} \label{Pelb}
\sqrt{P_e} \geq \frac{\ln(2)\bracs{H(K) -\ln(E+1) - 1}}{\ln\bracs{2(M-1)}},
\end{align}
which represents a fundamental tradeoff between $P_e$, $E$ and $M$. 

In particular, consider the possibility of \emph{zero-error communication} ($P_e =0$) for a given $M$. For equiprobable messages, Eq.~\eqref{Pelb} implies that
\begin{align} \label{ncZE}
E \geq M/e - 1
\end{align} 
is a necessary condition for this to be possible. Sending the number states $\{\ket{0}, \ldots, \ket{M-1}\}$ and performing photon counting at the receiver achieves zero error with an average energy of $(M-1)/2$.  Transmitting the eigenstates of the Pegg-Barnett phase operator \cite{PB89} defined on a space truncated to a maximum photon number of $M-1$ also achieves zero error at the same average energy \cite{NYG+12}. Although the condition \eqref{ncZE} is not tight for these examples, it implies that no other $M$-ary communication scheme can achieve zero error with a substantially smaller average energy.

\section{Heisenberg limit for Multimode multipass phase estimation} \label{sec:mmmp}

In Section~\ref{sec:1modeestimation}, the performance of phase modulation schemes involving multiple signal modes, ancilla-assisted schemes, and schemes involving nonlinear modulation were reduced to that involving a single signal mode. However, in all those schemes, a given signal mode is modulated by the phase element only once. It may be feasible in some situations for each signal mode to be linearly modulated by the phase being sensed multiple times in so-called \emph{sequential schemes} \cite{GLM11}.  It is well-known in the  QFI-based approach that such sequential schemes may result in greater measurement precision for the same resources  -- see, e.g., \cite{CPR00,Lui02}. In this section, we derive limits on the CMSE of multimode multipass schemes for phase estimation under an energy constraint. Such schemes (see Fig.~4) are a hybrid of parallel and sequential strategies and have been studied both theoretically and experimentallly \cite{CPR00,HBB+07,HBB+09,XHB+11}.

For optical phase estimation in the absence of additional loss or noise, and in the Bayesian framework considered here, we showed in Sec.~\ref{sec:1modeestimation} that the most general ancilla-assisted entangled parallel strategy has the same performance as a suitable single-signal-mode ancilla-free strategy. For ancilla-free multipass, i.e., sequential strategies with a single signal mode, the analysis of Sec.~\ref{sec:1modeestimation} goes through unchanged and the Heisenberg scaling \eqref{Hlimit} with respect to the average signal energy cannot be beaten. 

\begin{figure}[tbp]
    \includegraphics[trim=74mm 75mm 75mm 75mm, clip=true,width=0.6\linewidth]{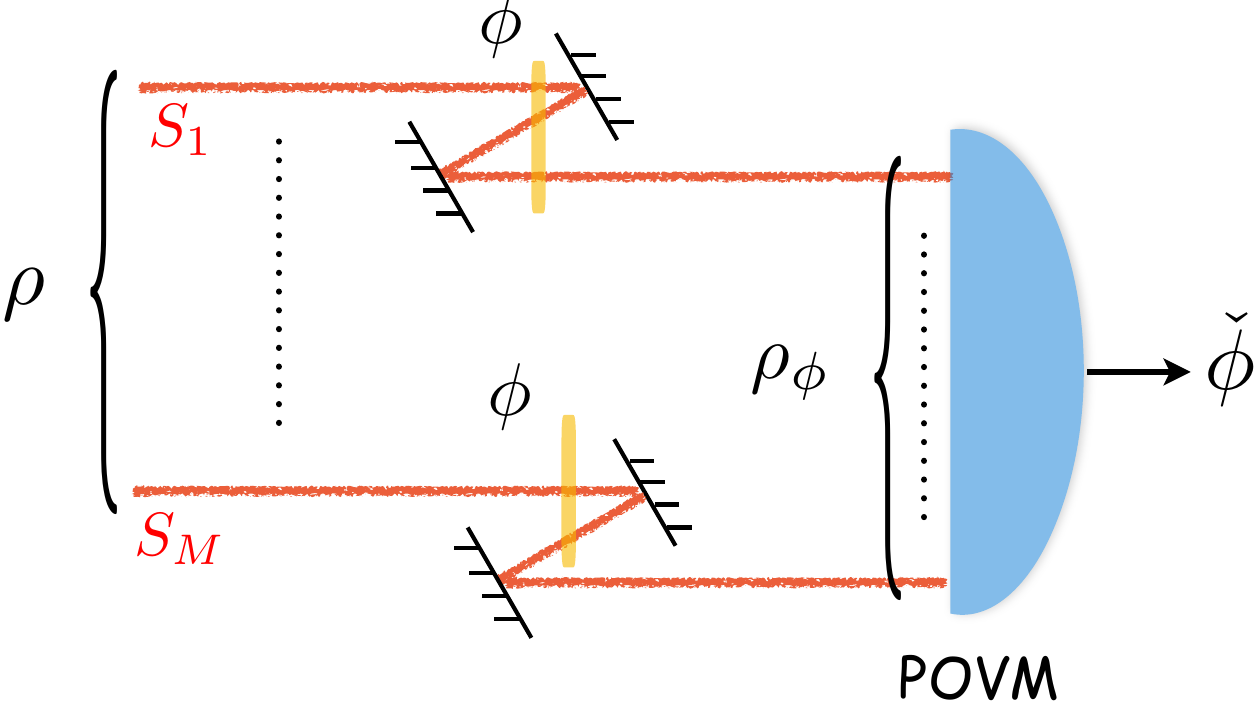}
    \caption{Schematic of a multimode multipass phase estimation setup: A probe state $\rho$ of $M$ signal modes $S_1,\ldots, S_M$ is used to estimate the phase $\phi$ of a phase element. The $m$-th mode makes $\pi_m$ passes through the phase element.}     \label{fig:multipass}
\end{figure}

Let us now consider multipass strategies using $M$ signal modes. The unrestricted capacity of $M$ optical modes with a total energy constraint of $E$ across the modes is $M\,g\pars{E/M}$ and applying the ITI yields a lower bound on the CMSE $\propto \exp(-E)$ in the limit $M \rightarrow \infty$, which -- if realizable -- is potentially highly significant \cite{Yue92,HW12b}.
We can leverage the results of \cite{HW12b} to get a tighter bound for multimode multipass phase estimation. Suppose that the $m$-th mode makes $\pi_m$ passes through the phase element (Fig.~\ref{fig:multipass}) -- see \cite{HBB+07} for an experimental realization that exploits the quantum phase estimation algorithm \cite{CPR00,NC00}. For $\hat{N}_m$ the number operator of the $m$-th mode,  the $M$-mode probe state $\rho$ is transformed as:
\begin{align}
\rho \mapsto \rho_\phi = \exp(-i\phi\hat{G})\, \rho \, \exp(i\phi\hat{G}),
\end{align}
with the generator $\hat{G}$ given by:
\begin{align}
\hat{G} = \sum_{m=1}^M \pi_m\,\hat{N}_m.
\end{align}
In \cite{HW12b}, it was shown that the mutual information (and hence the capacity under phase modulation) $I\pars{\Phi ; \check{\Phi}}$ between $\Phi$ and the outcome $\check{\Phi}$ of any POVM measured on the modulated state is bounded as
\begin{align}
I\pars{\Phi ; \check{\Phi}} \leq A_{\hat{G}}(\rho) \leq H\pars{\hat{G} | \rho}
\end{align}
where $A_{\hat{G}}(\rho)$ is the $\hat{G}$-asymmetry of the probe state \cite{VAW+08} which is in turn bounded by the Shannon entropy of the measurement of the observable $\hat{G}$ on the probe state. The spectrum of $\hat{G}$ consists of a subset of the non-negative integers so that
\begin{align}
H\pars{\hat{G} | \rho} \leq g\pars{\mean{\hat{G}}_{\rho}} =  g\pars{\sum_{m=1}^M \pi_m \mean{\hat{N}_m}_{\rho}} \leq g\pars{\pi_{\mr{max}} E},
\end{align}
where $\pi_{\mr{max}}$ is the largest of $\left\{\pi_m\right\}_{m=1}^M$. Together with the ITI, this yields the Heisenberg limit
\begin{align} 
{\rm{CMSE}} \geq \frac{Q_\Phi}{e^2} \frac {1}{(\pi_{\mr{max}}E+1)^2}. \label{mmmpHlimit}
\end{align} 
If the number of passes and the total energy are considered as independent parameters, Eq.~\eqref{mmmpHlimit} shows that Heisenberg scaling with the total energy $E$ cannot be beaten by a multimode multipass strategy with fixed $M$ and $\{\pi_m\}_{m=1}^M$. In particular, the exponential scaling with respect to the average number of photons demonstrated in \cite{HBB+07} can be seen as a consequence of the maximum number of passes in the setup being itself exponential in the number of photons. The bound \eqref{mmmpHlimit} suggests that the route of multimode multipass linear phase modulation is impractical for obtaining super-Heisenberg scaling with respect to the total average energy in the signal modes.

\section{Quantum limit for ancilla-assisted phase estimation in loss} \label{sec:lossyphaseest}

As our final example of the application of the rate-distortion approach to phase estimation, we consider phase estimation in the presence of the experimentally ubiquitous loss. As a first approach to the problem,  consider the single-mode scenario of Fig.~\ref{fig:1modephaseest} with the addition of a loss element in the optical beam post the phase modulation represented by a beamsplitter of transmittance $\eta <1$. The analysis of Sec.~\ref{sec:1modeestimation} can then be carried through as before provided only that the capacity $g(E)$ of the noiseless single-mode channel is replaced by that of a lossy single-mode channel, which is known to be $g(\eta E)$ \cite{GGL+04}. The resulting lower bound on the CMSE is therefore the same as  \eqref{Hlimit} with $E$ replaced by $\eta E$, which is a stronger bound but still shows Heisenberg scaling relative to $E$.

Unlike the noiseless single-mode channel, the only known ensemble attaining the capacity of the lossy single-mode channel consists of coherent states of mean amplitude distributed according to a circularly symmetric Gaussian distribution in phase space \cite{GGL+04}. Such an ensemble is generated by applying Gaussian-distributed displacements to the vacuum state. Thus, it is reasonable to expect that the capacity of the lossy channel with an input ensemble generated via phase modulation on a probe state of energy $E$ is strictly less than its unrestricted capacity $g(\eta E)$. We show below that this is indeed the case even for more general ancilla-assisted phase estimation schemes.

\begin{figure}[tbp]
    \includegraphics[trim=60mm 95mm 76mm 96mm, clip=true,scale=0.55]{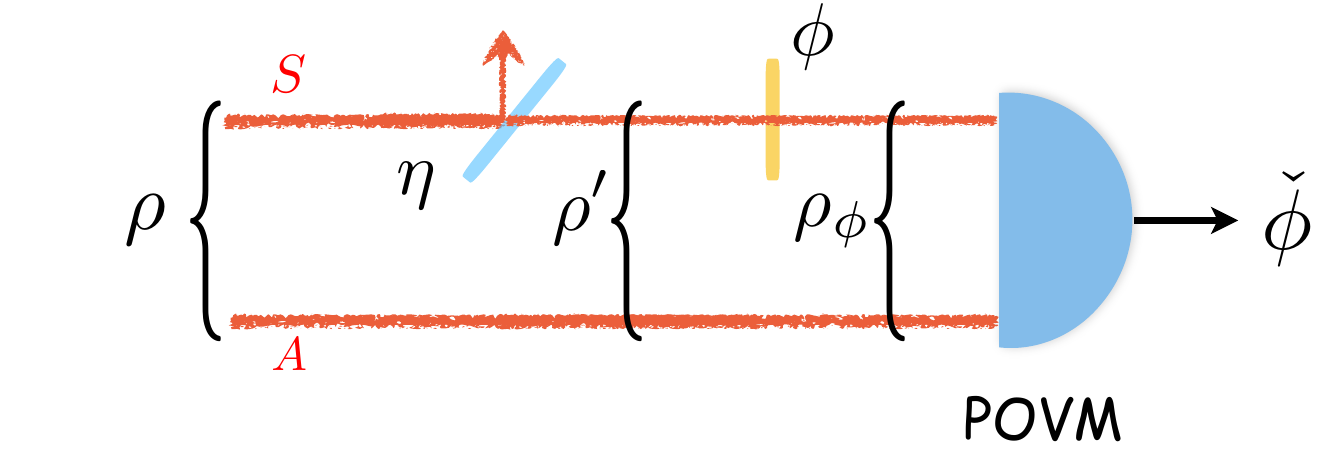}
    \caption{Estimation of a phase parameter $\phi$ using a signal-ancilla entangled probe $\rho$. The signal mode $S$ suffers a loss $1-\eta$ depicted as a beam splitter of transmittance $\eta <1$ while the ancilla mode $A$ is not degraded. The effective probe state $\rho'$ of Eq.~\eqref{rholoss} is also indicated.}     \label{fig:lossyphaseest}
\end{figure}

Consider the ancilla-assisted phase estimation scheme  shown in Fig.~\ref{fig:lossyphaseest} where the signal mode $S$ suffers nonzero loss represented by the beamsplitter of transmittance $\eta <1$\footnote{We note that the composition of loss and phase shift channels shown in  Fig.~\ref{fig:lossyphaseest} produces the correct output state $\rho_\phi$  independent of the order in which the loss interaction and phase shift are applied. The phase-shift and loss may also both be distributed over the path of the beam -- the output state depends only on the total values of these parameters.}. An average energy constraint of $E$ is imposed on the signal mode. We assume that the ancilla system $A$ (which can be arbitrary) is held noiselessly -- any loss or noise in this system can only worsen the performance, so that the bound we derive is valid regardless.We allow for arbitrary POVMs to be performed on the joint $SA$ system in order to obtain the phase estimate $\check{\Phi}$. 

An arbitrary pure-state probe on the joint signal-ancilla system can be written as
\begin{align} \label{NDSprobe}
\ket{\psi}  = \sum_{n=0}^\infty \sqrt{p_n} \kets{n} \keta{\xi_n},
\end{align}
where $\{\kets{n}\}$ are number states of the signal mode, $\{p_n\}$ is the probability distribution of the signal photon number and $\{\keta{\xi_n}\}$ are arbitrary normalized states of the ancilla. Due to our assumption that the ancilla system is undisturbed, the state of Eq.~\eqref{NDSprobe} can be isometrically mapped into \emph{any} other ancilla Hilbert space with countably infinite dimension and made to yield the same performance with an appropriately transformed POVM on the target Hilbert space. Without loss of generality, therefore, we can take the ancilla system to be a single bosonic mode as shown in Fig.~\ref{fig:lossyphaseest}. This conclusion is also valid for a mixed-state probe since it can first be purified into the form \eqref{NDSprobe} on a larger ancilla system before arguing as above. 

For any $\eta <1$, and for given $\{p_n\}$, it was shown in \cite{NY11} (Theorem 1 therein) that the states \eqref{NDSprobe} with the $\left\{\ket{\xi_n}_A\right\}$ taken to be mutually orthonormal optimize any Bayesian cost function, and therefore the CMSE as well. Such probes are called \emph{Number-Diagonal-Signal (NDS)} probes since the reduced density operator of the signal mode is diagonal in the number basis, and we can confine attention to them in order to derive a lower bound on the CMSE.

The state $\rho'$ of $SA$ after the loss interaction but before the phase modulation (see Fig.~\ref{fig:lossyphaseest}) can be calculated by standard techniques, e.g., by explicitly calculating the output state from the unitary interaction between the signal and an environment mode initially in the vacuum state followed by tracing out the environment mode.  If the probe state \eqref{NDSprobe} is NDS, it turns out that $\rho'$ has the spectral decomposition
\begin{align} \label{rholoss}
\rho' = \sum_{l=0}^\infty q_l\, \ket{\chi_l}\bra{\chi_l}.
\end{align}
Here, $q_l$ can be interpreted as the probability that $l$ photons are lost to the environment during the beam-splitter interaction  and is given by
\begin{align} \label{ql}
q_l =  \sum_{n \geq l} p_n \binom{n}{l}\, \eta^{n-l}\, (1-\eta)^l  \equiv   \sum_{n \geq l} p_n  B_\eta(n,l).
\end{align}
The states $\{\ket{\chi_l}\}$ in eq.~\eqref{rholoss} are given by
\begin{align} \label{Chil}
\ket{\chi_l} = \frac{1}{\sqrt{q_l}}\sum_{n \geq l}\sqrt{ p_n \, B_\eta(n,l)} \;\kets{n-l}\keta{\xi_n} 
\end{align}
and form an orthonormal set, i.e., $\braket{\chi_l}{\chi_{l'}}= \delta_{l,l'}$ by virtue of the fact that the $\{\keta{\xi_n}\}$ are orthonormal. The modulated state $\rho_\phi$ of Fig.~\ref{fig:lossyphaseest} is therefore given by 
\begin{align}
\rho_\phi = \sum_l q_l\, \ket{\chi_l(\phi)}\bra{\chi_l(\phi)},
\end{align}
where
\begin{align}
\ket{\chi_l(\phi)} = \frac{1}{\sqrt{q_l}}\sum_{n \geq l}\sqrt{ p_n \, B_\eta(n,l)} \;e^{in\phi}\, \kets{n-l}\keta{\xi_n}.
\end{align}

For $\overline{\rho} = \int d\phi \, P_{\Phi}(\phi) \,\rho_\phi$ the average modulated state, the Holevo bound on the mutual information between $\Phi$ and its estimate $\check{\Phi}$ reads
\begin{align}
I(\Phi; \check{\Phi})&\leq S(\overline{\rho}) - \int_0^{2\pi} d\phi \,P_\Phi(\phi) \,S(\rho_\phi), \\
&= S(\overline{\rho}) - S(\rho') \\
&\leq S\pars{\cl{N}\pars{\overline{\rho}}} - S(\rho') \\
&= S\pars{\cl{N}\pars{\rho'}} - S(\rho')
\end{align}
where  $\cl{N}$ is the unital quantum channel of Eq.~\eqref{Ndef} so that 
\begin{align} \label{Mrho'}
\cl{N}\pars{\rho'} = \sum_{l=0}^\infty \sum_{n=l}^\infty p_n \,B_\eta(n,l)\,\kets{n-l}\bra{n-l} \otimes  \keta{\xi_n}\bra{\xi_n}.
\end{align}
The orthogonality of $\{\keta{\xi_n}\}$ then implies that
\begin{align}
S(\cl{N}\pars{\rho'}) = H(N,N-L),
\end{align}
where $N$ and $N-L$ are the classical random variables corresponding to a measurement on $\cl{N}\pars{\rho'}$ of the index of the $\{\keta{\xi_n}\}$ basis on the ancilla mode and the photon number in the signal mode respectively, and $H(\cdot)$ is the Shannon entropy. Similarly, the orthogonality of  $\{\ket{\chi_l}\}$ implies that
\begin{align}
S(\rho')= H(L).
\end{align}
Combining the above facts, we have
\begin{align}
I\pars{\Phi; \check{\Phi}} &\leq S(\cl{N}\pars{\rho'}) - S(\rho') \\
&= H(N,N-L) - H(L) \nonumber\\&= H(N,L) -H(L) \nonumber\\
&= H(L|N) - \left[H(L)-H(N)\right] \nonumber\\
& = \sum_n p_n\, H(L|N=n)  - \left[H(L)-H(N)\right] \label{CphUBa} \\ 
& \leq  \sum_n \frac{p_n}{2} \ln \left[ 2\pi e \left(\eta\,(1-\eta)\,n + \frac{1}{12}\right)\right] \nonumber\\&\hspace{2cm} - \left[H(L)-H(N)\right] \label{Diffentbound}\\
& \leq \frac{1}{2} \ln \left[ 2\pi e \left(\eta \,(1-\eta)\,E + \frac{1}{12}\right)\right] \nonumber\\& \hspace{2cm}- \left[H(L)-H(N)\right]. \label{CphUBb}
\end{align}
Here we have used standard entropy manipulations to obtain eq.~\eqref{CphUBa}. To obtain \eqref{Diffentbound}, we have used \eqref{diffentbound} and the fact that, conditioned on $N=n$, $L$ has the binomial distribution Bin$(n, 1-\eta)$. Eq.~\eqref{CphUBb} follows from concavity of the logarithm.

\begin{figure}[tbp]
    \includegraphics[trim=100mm 100mm 100mm 105mm, clip=true,width=0.6\columnwidth]{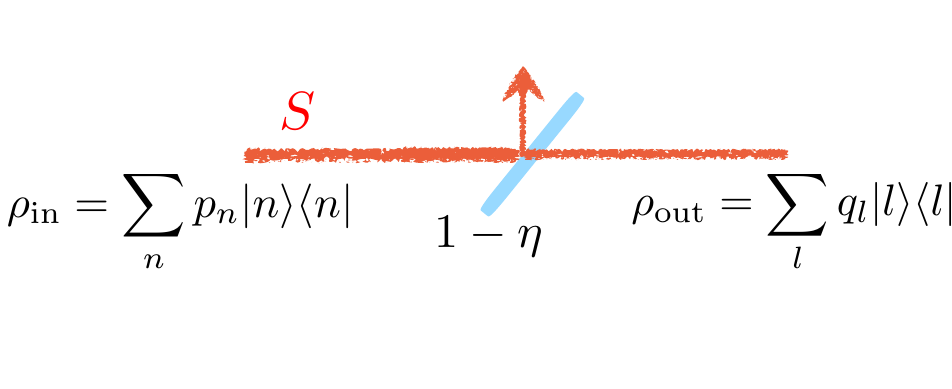}
    \caption{The single-mode state and loss channel of transmittance $1-\eta$ for which the entropy gain from input to output equals $H(L)-H(N)$ of Eq.~\eqref{CphUBb}.}       \label{fig:entropygain}
\end{figure}

We now bound the second term in eq.~\eqref{CphUBb}. Consider the single-mode pure-loss channel $\cl{L}$ depicted in Fig.~\ref{fig:entropygain}.  It is readily verified that, for the input state $\rho_{\rm in} = \sum_{n} p_n \ket{n}\bra{n}$, the channel outputs the state $\rho_{\rm out} = \sum_{l} q_l \ket{l}\bra{l}$ so that the \emph{entropy gain} from input to output is precisely $H(L)-H(N)$. For the channel $\mathcal{L}$ of Fig.~\ref{fig:entropygain}, Holevo has shown (See Theorem 2 of \cite{Hol10,*Hol10arxiv}) that the \emph{minimum} entropy gain
\begin{align}\inf_{\rho_{\rm in} \in \cl{H}_S }\left[S(\cl{L}\pars{\rho_{\rm in}})-S(\rho_{\rm in})\right] = \ln\left(1-\eta\right),
\end{align}
where the infimum is over all input states in $\cl{H}_{S}$ and therefore includes the input state of Fig.~\ref{fig:entropygain}. We thus have
\begin{align} \label{EntDiffLB}
\ln\, (1-\eta) \leq H(L) - H(N),
\end{align}
which, combined with \eqref{CphUBb}, gives the upper bound
\begin{align}
I\pars{\Phi; \check{\Phi}}\leq & \frac{1}{2} \ln \left[ \frac{2\pi e} {(1-\eta)^2} \left(\eta \,(1-\eta)\,E + \frac{1}{12}\right)\right]
\end{align}
on the mutual information. The now familiar argument using the ITI along with the Shannon lower bound \eqref{DRSLB} gives the sought lower bound
\begin{align} \label{LossyMSELB}
\mbox{CMSE} &\geq  \frac{Q_\Phi(1-\eta)^2}{ 2\pi e\,  \left[\eta \,(1-\eta)\,E + \frac{1}{12}\right]}
\end{align}
on the CMSE of lossy ancilla-assisted phase estimation. Observe that it exhibits SQL scaling in the energy for any $\eta <1$. For large $E$, it differs from the lower bound \eqref{CSbound} for coherent states (with $E$ replaced by $\eta E$ to account for the attenuation of the coherent-state amplitude due to the loss) by only a factor of $(1-\eta)$. We have thus obtained a very strong limitation on the improvement in CMSE that an ancilla-assisted phase estimation scheme using arbitrary nonclassical probe states can offer over the practically much simpler scheme using a single-mode laser-light probe.

\section{Discussion and Outlook} \label{sec:disc}
We have developed fundamental lower bounds on various phase estimation and communication scenarios involving one or more optical modes using rate-distortion theory. The bounds are valid for any prior probability distribution and arbitrary quantum measurements, including adaptive ones. The bounds explicitly display dependencies on the system design parameters and energy constraints, as well as on the prior statistics of the parameter via the entropy power of the  distribution. They are non-asymptotic and valid for all values of the average energy $E$ including in the limit of $E \rightarrow 0$ in which the minimum MSE (or CMSE) is dictated purely by the amount of prior information available. For phase estimation, our bounds are given in terms of the stronger CMSE criterion naturally suited to phase parameters.

The Heisenberg limits obtained in Sec.~\ref{sec:1modeestimation} for single-mode linear and nonlinear phase estimation schemes agree with those in ref.~\cite{HW12b}, while that for the multimode multipass scheme of Sec.~\ref{sec:mmmp} is stronger than that in refs.~\cite{Yue92,HW12b}.
We should mention that  SQL scaling for phase estimation (with the usual linear modulation) in the presence of loss has also been derived in the QFI-based quantum metrology framework \cite{EdMFD11,DKG12,Tsa13}. Thus, the following bound  for lossy phase estimation was shown in \cite{EdMFD11}:
\begin{align} \label{EdMFDLB}
\delta \Phi^2_\phi \geq \frac{1-\eta}{4\,\eta \,E} + \frac{1}{4\, \langle \Delta \hat{N}_S^2 \rangle_\rho},
\end{align}
where the left-hand side $\delta \Phi^2_\phi $ is the mean squared error achieved by any unbiased estimator for a particular (but arbitrary) value of $\phi$ and $\langle \Delta \hat{N}_S^2 \rangle_\rho$ is the variance of the signal photon number of the probe state $\rho$. This bound, based as it is on
the quantum Cram\'{e}r-Rao bound, regards $\phi$ as an unknown rather than random parameter and holds provided the estimate $\check{\Phi}$ is unbiased. In the regime of large $\langle \Delta \hat{N}_S^2 \rangle_\rho$ for a fixed $E$, the second term can be neglected and the $\eta$-dependence of the bounds \eqref{EdMFDLB} and \eqref{LossyMSELB} is rather similar. It should be noted, however, that the latter bound holds for biased measurements as well and does not diverge in the limit $E \rightarrow 0$. Some frequentist bounds on the mean squared error of phase estimation explicitly depend on the parameter $\phi$ -- see, e.g., ref.~\cite{LJW13} -- indicating greater phase sensitivity at some points. We leave for future work the question whether Bayesian bounds tighter than those developed here can display such phase dependence, perhaps using tighter bounds on the phase-modulation capacity. A Bayesian bound showing SQL scaling for joint signal-ancilla states of fixed total photon number and a uniform phase prior has also been derived in \cite{KD-D10}.

The results obtained here can potentially be extended in several directions. Natural multimode problems that may be tackled using the approach of this paper are optical ranging \cite{GLM01}, measuring transverse displacements of optical beams \cite{FFM00,DTF+08}, and general image estimation problems \cite{NY11}. The technique may also be extended to multiple-parameter problems \cite{Hal18}, and even to the estimation of continuous waveforms \cite{BTH+15} building on the rate-distortion theory for sources generating random processes \cite{Ber03rd}. The results of Sec.~\ref{sec:lossyphaseest} appear to be extendable to multimode lossy scenarios. It would also be interesting to attempt to generalize them to multiple-parameter scenarios with other kinds of decoherence such as phase diffusion \cite{CDB+14,VDG+14,SBD17}.

\section{Acknowledgments}
Valuable discussions with Rafa{\l} Demkowicz-Dobrza{\'n}ski,  Saikat Guha, Michael J. W.~Hall, Marcin Jarzyna, Jan Ko{\l}ody{\'n}ski, Mankei Tsang, and Brent J.~Yen are much appreciated. I am grateful to Horace P.~Yuen for introducing me to the rate-distortion approach in quantum metrology. This work is supported by the Singapore National Research Foundation under NRF Grant No.~NRF-NRFF2011-07 and the Singapore Ministry of Education Academic Research Fund Tier 1 Project R-263-000-C06-112.

 %%%%%%%%%%%%%%%%%%%%%
%merlin.mbs apsrev4-1.bst 2010-07-25 4.21a (PWD, AO, DPC) hacked
%Control: key (0)
%Control: author (72) initials jnrlst
%Control: editor formatted (1) identically to author
%Control: production of article title (-1) disabled
%Control: page (0) single
%Control: year (1) truncated
%Control: production of eprint (0) enabled
%

\end{document}